\documentstyle[graphicx,amsmath,amssymb,12pt]{article}
\begin{document}
\setlength{\parskip}{0.45cm}
\setlength{\baselineskip}{0.75cm}
%XXXXXXXXXXXXXXXXXXXXXXXXXXXXXXXXXXXXXX
%
%SETTINGS FOR PREPRINT-SPACED VERSION
%setlength{\parskip}{0.45cm}
%setlength{\baselineskip}{0.75cm}
%
% SETTINGS FOR DOUBLE - SPACED VERSION
%\setlength{\parskip}{0.65cm}
%\setlength{\baselineskip}{0.95cm}
%
%XXXXXXXXXXXXXXXXXXXXXXXXXXXXXXXXXXXXXX
\begin{titlepage}
\setlength{\parskip}{0.25cm}
\setlength{\baselineskip}{0.25cm}
\begin{flushright}
DO-TH 03/01\\
hep--ph/\\
\vspace{0.2cm}
%hep--ph/0103137\\
%\vspace{0.2cm}
%November 2001
\end{flushright}
\vspace{1.0cm}
\begin{center}
\LARGE
{\bf NEUTRINO PRODUCTION OF RESONANCES}

\vspace{1.5cm}
\large
Emmanuel A.\ Paschos and Ji--Young\ Yu\\ \vspace{1.0cm}  
\normalsize
{\it Universit\"{a}t Dortmund, Institut f\"{u}r Physik,}\\
{\it D-44221 Dortmund, Germany} \\
\vspace{0.5cm}
\large{and}
\vspace{0.5cm}
\large
Makoto Sakuda\\ \vspace{1.0cm}
\normalsize
{\it High Energy Accelerator Research Organization (KEK)}\\
{\it Tsukuba 305-0801, Japan}\\
\vspace{1.5cm}
\end{center}

\begin{abstract}

We take a fresh look at the analysis of resonance production by neutrinos.
We consider three resonances $P_{33}, P_{11}$ and $S_{11}$
with a detailed discussion of their form factors. 
The article presents results for free proton and 
neutron targets and discusses the corrections 
which appear on nuclear targets.  
The Pauli suppression factor is derived in the 
Fermi gas model and  shown to apply to resonance production. 
The importance of the various resonances is demonstrated 
with numerical calculations. The $\Delta$-resonance is described by two 
formfactors and its differential cross sections are 
compared with experimental data.
The article is self-contained and could be helpful to readers who wish to  
reproduce and use these cross sections.
\end{abstract}
\end{titlepage}

%MAIN PART
\renewcommand{\theequation}{\arabic{section}.\arabic{equation}}
\section{Introduction}

The excitation of the resonances by electrons and
neutrinos has been studied for a long time.  The
earlier articles \cite{ref1}--\cite{ref5} tried to determine
the $p\Delta$--transition form factors in terms of 
basic principles, like CVC, PCAC, dispersion 
relations, etc.  These and subsequent papers 
introduced dipole form factors and in various cases
other functional forms with
additional kinematic factors in order to reproduce the data. 
The result was the presentation of cross sections
(differential and integrated) in terms of several 
parameters \cite{ref6}--\cite{ref9}.  
The relatively large number of parameters 
and the limited statistics of the experiments 
provided qualitative comparisons but an accurate
determination of the terms is still missing.
A new generation of experiments is now under
construction aiming to measure properties of the
neutrino oscillations and they present the 
opportunity for precise tests of the standard
model.

For these reasons we decided to improve the 
calculation of the excitation of resonances with
isospin $I=3/2$ and $I=1/2$ looking into the 
various terms that enter the calculations and
trying to determine them, as accurately as possible.
For reasons of economy we shall study four 
resonances. We shall give explicit formulas and form factors for  
$P_{33}(1232)$, $P_{11}(1440)$,
and $S_{11}(1535)$ and use a functional form for $D_{13}(1520)$ 
obtained in electroproduction \cite{ref11}.
We calculate the cross section for the production of 
each resonance alone (fig. 3) and found
that $P_{33}(1232)$ dominates.
Next we shall study the
vector current in electroproduction data and
then use it for neutrino reactions.  For the 
axial current we shall discuss the constraints
introduced by PCAC for several form factors.
Then we shall calculate the differential cross
section in $Q^2$ and also in $W$ (invariant mass
of the final pion--nucleon system).  
We shall restrict the numerical analysis to the $p\pi^+$ channel 
and to $W {\stackrel{<}{\sim}} \,1.6$ GeV where 
there is accurate data and in order to avoid the 
influence of higher resonances. For example the $P_{33} (1640)$ 
resonance is expected to provide a small contribution, at low energies, 
because it is further away and has a small elasticity \cite{ref8}.
For our kinematic region the higher resonances with $I = 1/2$ 
provide a smooth background to the $p\pi^0$ and $n\pi^+$ channels.
Their effects will show up in the W-distributions and the 
overall scale of the integrated cross sections.
This approach was demonstrated 
\cite{ref10} to agree with the available data.
In this article we extend the calculations 
giving more details which will be helpful for
future comparisons.  Cross sections will be
presented primarily for free protons and 
neutrons.  
In the future we will extent this study by 
including higher resonances contributing to other channels.
It will be interesting to include the inelasticity of higher resonance 
in order to estimate their contribution to the multi-pion events.

Neutrino experiments, however, use medium--heavy and 
heavy nuclei which brings additional corrections.
Several studies so far identified nuclear effects
from
\begin{itemize}
\item[i)] the Pauli exclusion principle,
\item[ii)] the Fermi motion, and 
\item[iii)] the absorption and charge exchange of the produced pions
\end{itemize}
in nuclei.  In this article we discuss in greater
detail the Pauli effect and show that it brings
a modification to the data which should be 
identified and checked in the experiments.
It may be important in producing the $Q^2$ distribution.

We shall adopt a notation similar to the one in
deep inelastic scattering 
\begin{eqnarray}
\nu(k)+p(p)\to \mu^-(k') 
& + & \Delta^{++}(p')\nonumber\\
& & \hookrightarrow p(p'') +\pi(p_{\pi}).
\end{eqnarray}
The momentum transfer will be denoted by $q=k-k'$,  
which is space--like with
\begin{equation}
Q^2 = - q^2
    = - m_{\mu}^2+4E \left(E'-\left[E'^2
      - m_{\mu}^2\right]^{1/2}\cos\theta\right).
\end{equation}
The energy of the current in the laboratory and 
in the rest frame of the resonance, to be denoted
by CM, is given by
\begin{eqnarray}
q_L^0 =  \nu   & = & \frac{q\cdot p}{M}= \frac{W^2+Q^2-M^2}{2M}\nonumber\\
q_{\rm CM}^0 & = & \frac{W^2-Q^2-M^2}{2W}
\quad\quad{\rm and}\quad\quad 
q_{\rm CM} = \left[ \left( q_{\rm CM}^0\right)^2 +Q^2\right]^{1/2}.
\end{eqnarray}
Finally, we give the energy of the pion in
the rest frame of the resonance
\begin{equation}
p_{\pi}^0(W) = \frac{W^2+m_{\pi}^2-M^2}{2W} .
\end{equation}

We shall try to give explicit formulas for the 
cross sections so that the interested reader
will be able to reproduce the results.

The paper is planned as follows.
In section 2 we present the general formalism for the production of the 
$\Delta$-resonance emphasizing the minimal input, which is necessary.
In section 3 we give explicit formulas for the other two resonances $P_{11}$ 
and $S_{11}$ where contribution we find to be small.

Since nuclear corrections were calculated long time ago we decided to present 
an explicit calculation of the Pauli suppression factor, 
in order to examine its validity in various processes.
We emphasize in section 4 how the Pauli factor should be 
applied to quasi-elastic scattering, 
as well as in the production of resonances.
In section 5 we apply the formalism to the production 
of $\Delta^{++}$ and rely on the connection with electroproduction data.
Finally, in section 6 is presented a summary of the results.

%Section 2
\renewcommand{\theequation}{\arabic{section}.\arabic{equation}}
\setcounter{equation}{0}
\section{Cross Sections and their Form Factors}

Photoproduction and electroproduction data in the $\Delta$-region 
are well reproduced by a single 
form factor plus a smooth background representing the 
tails of higher resonances \cite{ref11}.  We shall use these results
by adopting one vector form factor, i.e. the magnetic
dipole term. 

For neutrino--induced reactions, on the other hand, there 
are many models some of which introduce many resonances.
This way one accounts for the data at the cost of 
introducing many parameters.
Other models restrict the analysis in the $\Delta$--region,
$W$ \raisebox{-0.1cm}{$\stackrel{<}{\sim}$} 1.4 GeV,
where the $\Delta$--resonance dominates and the main 
issue is the selection of the axial form factors and 
their parameterization. 
We shall use form factors which
satisfy PCAC and have a modified dipole $Q^2$-- dependence.  This 
approach provides a relatively simple formalism for single
pion production and was shown recently to 
reproduce the existing 
data \cite{ref10}.  We shall adopt this formalism and 
use it to calculate differential and total cross sections
comparing it with data in order to
determine overall validity or need for modifications and
extensions. 
In this article, the formulas for the excitation of the $\Delta$ resonance 
are simpler than in ref. \cite{ref5} because we take into account 
special properties of the form factors, 
which have been accumulated in the meanwhile. We give explicit 
formulas for the $I = 1/2$ resonances and show 
that in the energy region of the experiments their contribution is small. 

For the matrix element of the vector current we use
the general form
\begin{equation}
\langle \Delta^{++}|V_{\mu}|p\rangle =
\left[ \bar{\psi}_{\mu} A_{\lambda}q^{\lambda}-
\bar{\psi}_{\lambda}q^{\lambda} A_{\mu}+C_6^V(q^2)
\bar{\psi}_{\mu}\right] \gamma_5 \,u(p)\,f(W)
\end{equation}
with $A_{\lambda}=\frac{C_3^V}{M}\gamma_{\lambda}
+\frac{C_4^V}{M^2} p'_{\lambda}+\frac{C_5^V}{M^2}p_{\lambda}$,
$M$ the proton mass and $C_i^V$, $i=3, \ldots, 6$
the vector form factor, $\psi_{\mu}$ is the 
Rarita--Schwinger wave function of the $\Delta$--resonance
and $f(W)$ is the $s$--wave Breit--Wigner resonance,
given explicitly in eq.\ (2.5).
The conservation of the vector current (CVC) gives
$C_6^V(q^2)=0$ and the other form factors are determined
from electroproduction experiments, where the magnetic
form factor dominates.  This dominance leads to the 
conditions
\begin{equation}
C_4^V(q^2) = -\frac{M}{W}C_3^V(q^2)\quad\quad
              {\rm{and}} \quad\quad C_5^V(q^2) 
           = 0.
\end{equation}
With these conditions the electroproduction data depend
only on one vector form factor $C_3^V(q^2)$.
Precise electroproduction data determined the form
factor, which can be parametrized in
various forms.

Early articles, describe the static theory \cite{ref12} and the quark model
\cite{ref13} predicting the form factor for 
the $\gamma N \Delta$ vertex to be proportional to the isovector part of 
the nucleon form factors. Subsequent data \cite{ref11} 
showed that the form factor for $\Delta$-electroproduction 
falls faster with increasing $Q^2$ than the nucleon form factor 
which motivated some authors to introduce 
other parameterizations including exponentials \cite{ref14} 
and modified dipole \cite{ref15}.
The functional form
\begin{equation}
C_3^V(Q^2) = \frac{C_3^V(0)}{\left[1+\frac{Q^2}{M_V^2}\right]^2}
           \left(\frac{1}{1+\frac{Q^2}{4 M_V^2}}\right)
\end{equation}
gives an accurate representation. 
In this article we adopt this vector form factor 
and use CVC to determine its contribution to the neutrino induced reactions.
Details of the vector and axial contributions are presented in section five, 
where we 
shall estimate the contribution of $C_3^V$ from the electroproduction data.

The matrix element of the axial current has a similar
parameterization
\begin{equation}
\langle \Delta^{++}|A_{\mu}|p\rangle = 
  \left[\bar{\psi}_{\mu}B_{\lambda}q^{\lambda}-
  \bar{\psi}_{\lambda}q^{\lambda}B_{\mu}+
  \bar{\psi}_{\mu}C_5^A+\bar{\psi}_{\lambda}
  q^{\lambda}q_{\mu}C_6^A\right] u(p)f(W)
\end{equation}
with $B_{\lambda}=\frac{C_3^A}{M}\gamma_{\lambda}+
\frac{C_4^A}{M^2}p'_{\lambda}$ and
$\Gamma(W) = \Gamma_0 \frac{q_{\pi}(W)}{q_{\pi}(W_R)}$
and 
\begin{eqnarray}
f(W)=\frac{\sqrt{(\Gamma(W)/2\pi)}}{(W_R-W)-
\frac{1}{2} i\Gamma(W)} 
\end{eqnarray}
and \mbox{$\Gamma_0=120$ MeV}.

The PCAC condition gives the relation
$C_5^A(q^2) = -\frac{C_6^A}{M^2}q^2$
which for small $q^2=0$ leads to the numerical value
$C_5^A(0) = 1.2$ \cite{ref5}.
The contribution of the form factor $C_6^A$ to the 
cross section is proportional to the lepton mass 
and will be ignored. 

The $Q^2$--dependence of the form factors varies among
the publications giving different cross sections and
different $Q^2$  distributions even when the same $M_A$
is used.  For this dependence we shall use a modified dipole form
\begin{equation}
C_5^A(Q^2) = \frac{1.2}{\left[1+\frac{Q^2}{M_A^2}\right]^2} \Big
(\frac{1}{1+\frac{Q^2}{3 M_A^2}}\Big)\ .
\end{equation}
%eliminating an additional factor
%\begin{equation}
%\left( 1+\frac{aQ^2}{(b+Q^2)} \right)
%\end{equation}
%with $a=-1.21$ and $b=2$ GeV$^2$ 
%used in earlier publications \cite{ref5}. 
The proton has a charge distribution reflected in the form factor.
To build the resonance we must add a pion to the proton which creates 
a bound state with larger physical extent.
If the overlap of the wave functions has a larger mean-square-radius 
then the form factor will have a steeper $Q^2$ dependence 
as is indicated by the electromagnetic form factor for 
the excitation of $\Delta$ \cite{ref11}.
Since the effect is geometrical we expect a similar 
behavior for the vector and axial vector form factors. 
For this reason we replace another factor used in previous 
articles \cite{ref5} by the modified dipole in eq. (2.6) with 
$3 M_A^2 \sim 4 M_V^2 $.
For the other two form
factors $C_3^A(Q^2)$ and $C_4^A(Q^2)$ we shall use 
$C_3^A=0$ and $C_4^A(Q^2)=-\frac{1}{4}C_5^A$ \cite{ref5}.
It is evident that there is still arbitrariness
in the form factors  with $C_3^A$ and $C_4^A$ being small.
We show fig. \ref{forf} the relative contributions of the various terms, 
where it is clear that the contributions 
of $C_4^A$ and $C_4^V$ are indeed very small.

The differential cross section is finally given by
\begin{equation}
\frac{d\sigma}{dQ^2dW} 
= \frac{G^2}{16\pi M^2}\Big(\sum_{i=1}^3 K_iW_i\Big) g(Q^2,W)
\end{equation}
with the kinematic factors $K_i$ and the structure
functions $W_i(Q^2,W)$ defined in ref.\ \cite{ref5}. 
For our simplified case we collected the relevant
formulas and kinematics in Appendix A, so that the 
article is self--contained.  Most of the neutrino 
data give the integrated cross section
as function of the neutrino energy.  There are 
recent compilations of data \cite{ref10,ref16} which
have been compared with a theoretical calculation
\cite{ref10}.  Differential cross sections 
$\frac{d\sigma}{dQ^2}$ or $\frac{d\sigma}{dQ^2dW}$
were also reported by several experiments.  
In a high statistics
experiment at Brookhaven Laboratory, neutrino--Deuterium 
interactions \cite{ref17} were measured in the bubble
chamber.  Their data are precise 
and lead to the differential cross section shown
in figure \ref{bnl}.  The cross section falls off with 
increasing $Q^2$ and there is a dip at 
$Q^2\leq 0.2$ GeV. We shall investigate the distributions 
which are available in the simplified model described above.

%Section 3
\setcounter{equation}{0}

\section{Higher Resonances}

In addition to the $\Delta$-contribution 
there are higher mass resonances
$P_{11}(1440)$ and $S_{11}(1535)$ 
which contribute to the background
under $\Delta$ and, of course, at higher invariant masses. 
It is interesting to estimate 
their contribution as a function of neutrino energy, 
momentum-transfer-squared and invariant mass of the final state.
These are new resonances whose form factors are not known precisely.
We know accurately masses and 
widths of the resonances which are important for the calculations.

For the $P_{11}$ resonance we introduce the amplitude 
\begin{equation}
{\cal M}_P = \frac{G_F}{\sqrt{2}} \cos\theta_c\, l^\mu \, \bar{u}(p\prime)
             \gamma_5(\rlap/p + \rlap/q+M_R)\gamma_{\mu}
             (g_V - g_A \gamma_5) u(p)\, g_P\, f_P(W), 
\end{equation}
where $g_P$ denotes the coupling at the pion-nucleon resonance vertex 
and the Breit-Wigner factor
\begin{equation}
f_R(w) = \frac{1}{(W^2-M_R^2)+iM_R\Gamma(W)}
\end{equation} 
with 
$\Gamma_{R}(W) = \Gamma_{R}^0 \Big(\frac{q_\pi (W)}{q_\pi (M_{R})}\Big)^3$
and 
${q_\pi (W)} = \frac{1}{2 W} \big[(W^2-M^2-m_\pi^2)^2-4 M^2 m_\pi^2 \big]^{1/2}$ in the pion momentum in the rest frame of the resonance 
and $l^\mu$ is the leptonic tensor.  
The other form factors $g_V$ and $g_A$ are defined below.

Similarly, the amplitude for the production of $S_{11}$ is
\begin{equation}
{\cal M}_S = \frac{G_F}{\sqrt{2}} \cos\theta_c \, l^\mu \,\bar{u}(p\prime)
             (\rlap/p + \rlap/q+M_R)\gamma_{\mu}
             (g_V^{\prime} - g_A^{\prime} \gamma_5) u(p)\, g_S \,f_S(W), 
\end{equation}
the resonance factor $f_S(W)$ and the form factors 
$g_V^\prime$ and $g_A^\prime$ being now appropriate 
for the $S_{11}$ resonance. The functional form of the width 
is the same as for the $\Delta$-resonance:
$\Gamma_{S}(W) = \Gamma_{S}^0 \Big(\frac{q_\pi (W)}{q_\pi (M_{S})}\Big)$.  

The calculation of the new resonances requires knowledge 
of the form factors which are not known accurately. 
Usually, the form factors are obtained from model calculations.

Whenever we use $I=1/2$ resonances, we use the form factors \cite{ref6}:  
\begin{eqnarray}
g_P^V(Q^2) &=& 0\quad ,\\
g_S^V(Q^2) &=& -\frac{Q^2\, g^{2V}(Q^2)}{M_N (M_N -M_R)}\quad {\rm with} \\
g^{2V}(Q^2)& =& 
\frac{g^V(0)}{(1+\frac{Q^2}{4.3\, {\rm GeV^2}})^2}
\frac{1}{1+\frac{Q^2}{(M_R-M_N)^2}}\quad ,\\
g_R^A(Q^2) &=& \frac{g_R^A(0)}{(1+\frac{Q^2}{M_A})^2}
\end{eqnarray} 
The diffential cross section has the general form
\begin{equation}
\frac{d\sigma}{dQ^2dW} = 
\frac{G^2}{16\pi^3}\frac{q_\pi}{(k\cdot p)^2} 
\, V_R\,\frac{1}{(W^2-M_R^2)^2 + M_R^2\Gamma_R^2}\cdot g_R^2
\end{equation}
where the function $V_R$ given by 
\begin{eqnarray}
V_R &=& \cos^2\theta_c\,\Big[- 
            (g_V^2 - g_A^2)\, M_N\,  \big[2\, M_R\, p'\cdot\widetilde{k}
            \mp  M_N \,(W^2+M_R^2)\big] \,\frac{Q^2}{2} \nonumber\\
        &+& (g_V^2 + g_A^2) \big\{(p'\cdot\widetilde{k} 
            \mp  M_N\, M_R) \, M_N (Q^2 (E_f-E_i)+ 4 M_N E_i E_f) 
            \nonumber\\
        &+& (M_R^2-W^2) (M_N (\frac{Q^2}{2} (E_f-E_i) 
         +  2 E_i E_f (M_N-E_\pi)))\big\}\nonumber\\
        &+& g_V \,g_A \big\{2 (p'\cdot\widetilde{k} \mp M_N M_R) \,
            (\frac{1}{4} Q^2 (Q^2-2 M_N E_f))\nonumber\\
        &+& (M_R^2-W^2)\,(\frac{1}{4} Q^2 (Q^2- 2 M_N E_f+2 E_i E_\pi))
            \big\}\Big]
\end{eqnarray}
with $p'\cdot\widetilde{k} = W^2-W E_\pi$
and the upper and lower sign corresponding to the $P$ 
and $S$ resonances, respectively. 
The subscript $R = P,\,S$ signifies the $P_{11}$ and $S_{11}$ resonances.
We shall use these cross sections for calculating effects 
of higher resonances. 
We obtain  an overview of their importance by looking 
at the relevant contributions to the $W$- and $Q^2$-distributions. 
The results are shown for $E_\nu = 1.5 \, {\rm GeV}$ in figures \ref{dwres} (a) and 
\ref{dwres} (b).
The contribution of the $S_{11}$ resonance is everywhere very small.
The $P_{11}$ resonance gives a small contribution for $W>1.3 \,{\rm GeV}$ and may 
influence the $Q^2$-contribution.
The presence of a background at $W>1.3 \,{\rm GeV}$ shows up 
in the electroproduction data as well as indicated also in our fig. \ref{wu0}.
In fact in previous analysis of electroproduction of the
$\Delta$-resonance \cite{ref11} a polynomial dependence in $W$ 
was introduced to represent a background contribution.
The form factors and the other quantities, 
which we use, are summarized in Table \ref{tabl1}.
%table 1
\begin{table}[ht]
\par
\begin{center}
\vspace*{-0.5cm}
\begin{tabular}{|c|c|c|c|c|c|c|}
\hline
$ $ & $M_R\, ({\rm GeV})$ & $g_R$ & $\Gamma_R^0\, ({\rm GeV})$ & $g_R^V(0)$ & $g_R^A(0)$\\ \hline
$P_{11}$ & 1.440  & 4.45 & 0.35 &   0.0      & 0.35\\\hline
$S_{11}$ & 1.535  & 0.48 & 0.15 & - 0.28 (p) & 0.16\\
$      $ &        &      &      &   0.14 (n) &     \\\hline
\end{tabular} 
\end{center}
\caption{\sf Parameters} 
\label{tabl1}
\end{table}

\setcounter{equation}{0}

\section{The Pauli Effect}

Among the nuclear effects we shall describe in detail the 
Pauli suppression factor.
We shall assume the Fermi gas model with the nucleons enclosed
in a sphere with maximal momentum $p_F$. The Pauli 
exclusion principle requires the final wave
functions to be anti--symmetric in the exchange of two
identical particles.  
We assume the final wave functions to be
plane waves of the form
\begin{equation}
\psi(r_1,r_2) = e^{i(\vec{k_1}\vec{r_1}+\vec{k_2}\vec{r_2})}
              - e^{i(\vec{k_1}\vec{r_2}+\vec{k_2}\vec{r_1})}.
\end{equation}
The incoming current also brings a momentum $\vec{q}$ so
that the relevant matrix element is
\begin{equation}
{\cal{M}} = \int\psi^*(r_1,r_2)e^{-iq\cdot r_1}
             \varphi(r_1)\varphi(r_2)d^3{r_1}, d^3{r_2}
\end{equation}
where $\varphi(r_1)$ and $\varphi(r_2)$ are the wave 
functions of two bound nucleons.  We can regroup the terms and 
write the matrix element as
\begin{equation}
{\cal{M}} = F(k_1+q)F(k_2) - F(k_2+q)F(k1)
\end{equation}
with $F(k_1+q) = \int e^{-i(\vec{k_1}+\vec{q})\cdot\vec{r}}
 \varphi(r)dr$ and 
$F(k_1) = \int e^{-i\vec{k_1}\cdot\vec{r}}\varphi(r)dr$.
It is evident that the matrix element vanishes for 
$\vec{q}=0$, which we shall use as a condition later on.
The problem now is to carry out the integrals and express
the suppression factor in terms of $b = |\vec{q}|/p_F$, 
the ratio of momentum transfer to the Fermi momentum.

To obtain the cross section we square the amplitude and
integrate over final momenta
\begin{eqnarray}
\Sigma &=& \int|{\cal{M}}|^2 d^3{k_1} d^3{k_2} \nonumber\\
& = & \int|F(k_1+q)|^2d^3{k_1}\int|F(k_2)|^2d^3{k_2}
     + \int|F(k_1)|^2d^3{k_1}\int|F(k_2+q)|^2 d^3{k_2}\nonumber\\ 
& &  - 2\,Re \int F^*(k_1+q) F(k_1)d^3{k_1}\int F^*(k_2+q)F(k_2)d^3{k_2}. 
\end{eqnarray}
The terms $\int|F(k_2)|^2 \, d^3{k_2} =1$ involve the spectator
nucleon and will be normalized to one.  Similarly the 
integral $\int^{\infty} \big|F_1(k+q)\big|^2 \, d^3 k =1$ provided that the
integral falls very fast with $|\vec{k}+\vec{q}|^2\to\infty$.
This holds for the atomic form factors which are Gau\ss ian
or fall off exponentially for large momentum transfers.

The third integral in (3.4) is called the exchange term
and it is convenient to change the order of integration.
Performing the momentum integral first
\begin{equation}
I_0(r_1,r_2) = \int_{k_1=0}^{p_f} e^{i\vec{k_1}\cdot \vec{r_{21}}}d^3k_1
             = 4\pi\,\frac{p_f r_{21}\cos p_F r_{21}-\sin p_F r_{21}}
               {r_{21}^3}
\end{equation}
with $r_{21}=|\vec{r_2}-\vec{r_1}|$ and $p_f$ the
momentum of the Fermi sea.  The remaining integrations
over $r_1$ and $r_2$ are organized so that one of them is
over $(\vec{r_1}+\vec{r_2})$ giving a volume term and the
other is over $r_{21}=(\vec{r_2}-\vec{r_1})$.  The final
result is
\begin{equation}
\Sigma = |{\cal{M}}|^2 d^3{k_1} d^3{k_2} = 2-2V(2\pi)^4
                                  \int \cos(\vec{q}\cdot\vec{r_{21}})
                                  \frac{(p_F r_{21}\cos p_F r_{21}
                                - \sin p_F r_{21})^2}{r_{21}^6}\, d^3 r_{21}
\end{equation}
with the last integral being still over the 3--dimensional
$\vec{r}_{21}$ space.  The volume $V$ comes from the 
integration over 
$(\vec{r_1}+\vec{r_2})$ and is at this stage unspecified.
Defining $b=q/p_F$ and $z=p_F r_{21}$ the last integral
attains the form
\begin{equation}
\int_0^{\infty} \frac{\sin bz(\sin z-z\cos z)^2}{z^5}dz
= \frac{\pi}{24} \left[\frac{1}{4}b^4-3b^2+4b\right]
\end{equation}
for $0<b<2$ and equal to zero for $b=2$.  The evaluation
of this integral is not immediately evident and it can be
calculated with computer programs (Mathematica) or with 
the help of the residue theorem as 
described by Gatto \cite{ref18}.

\begin{figure}[htb]
\centering
\vspace*{-3.5cm}
\includegraphics[angle=270,width=15cm]{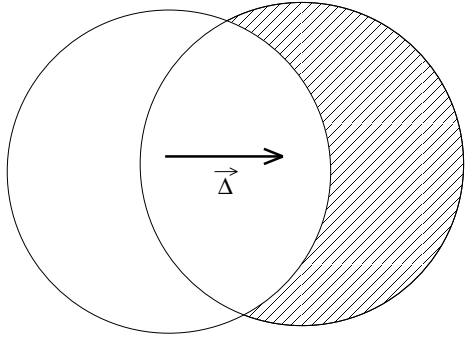}
%\subfigure{\epsfig{figure=pauli.eps,height=3.5in,angle=0}}
\vspace*{-2.5cm}
\caption{\sf Geometrical description of the Pauli suppression factor.}
\label{paulips}
\end{figure}

As we collect the various terms together, we must still
deal with the volume $V$ appearing in eq.\ (4.6).  We 
combine the volume together with other multiplicative
constants in a new constant $K$, so that the cross section
attains the form
\begin{eqnarray}
\Sigma & = & 2-2K p_F\frac{1}{b}(b^4-3b^2+4b)\\
& & = (2-8K p_F) + 2K p_F(3b-b^3).
\end{eqnarray}
Now we impose the condition that $\Sigma$ vanishes for
$b=0$.  This gives
\begin{equation}
2 K p_F = \frac{1}{2}\quad\quad{\rm and}\quad\quad
\Sigma = \frac{3}{2}b-\frac{1}{2} b^3\, .
\end{equation}
$\Sigma$ represents the fraction of the nucleons which 
can contribute for a given momentum transfer $q$.  It has
a geometrical interpretation frequently used in articles:
when a momentum $q$ is transferred to a nucleon, the center
of the Fermi sea is displaced by this momentum.  The 
fraction of the nucleons contributing to a cross section is the fraction of
the displaced sphere which lies outside the original
Fermi surface. The allowed region is the shaded volume
shown in figure \ref{paulips}.

The above derivation of the Pauli factor depends on the
approximation of treating the nucleus as a collection
of independent protons and neutrons.  The suppression
factor depends on the ratio of the momentum transfer
to the Fermi momentum and has a simple geometric
interpretation.  It does not depend on the specific
process and should hold for elastic and resonance
production, provided we calculate the ratio $b = q/p_F$
appropriate for each process.

For quasi--elastic scattering the Pauli factor has been
used in several articles and it was found to be necessary.
One assumes that there is one Fermi surface for protons
and neutrons for 
which the factor $\Sigma$ from (4.10) applies.  Some
authors assumed that there are distinct Fermi surfaces
for neutrons and protons and obtained two expressions
\cite{ref19,ref20}.  A detailed discussion for elastic
scattering is given by Berman \cite{ref19} and also
in the review article of Llewellyn--Smith \cite{ref4}.

\begin{figure}[htb]
\centering
\vspace*{-0.4cm}
\includegraphics[angle=270,width=15.cm]{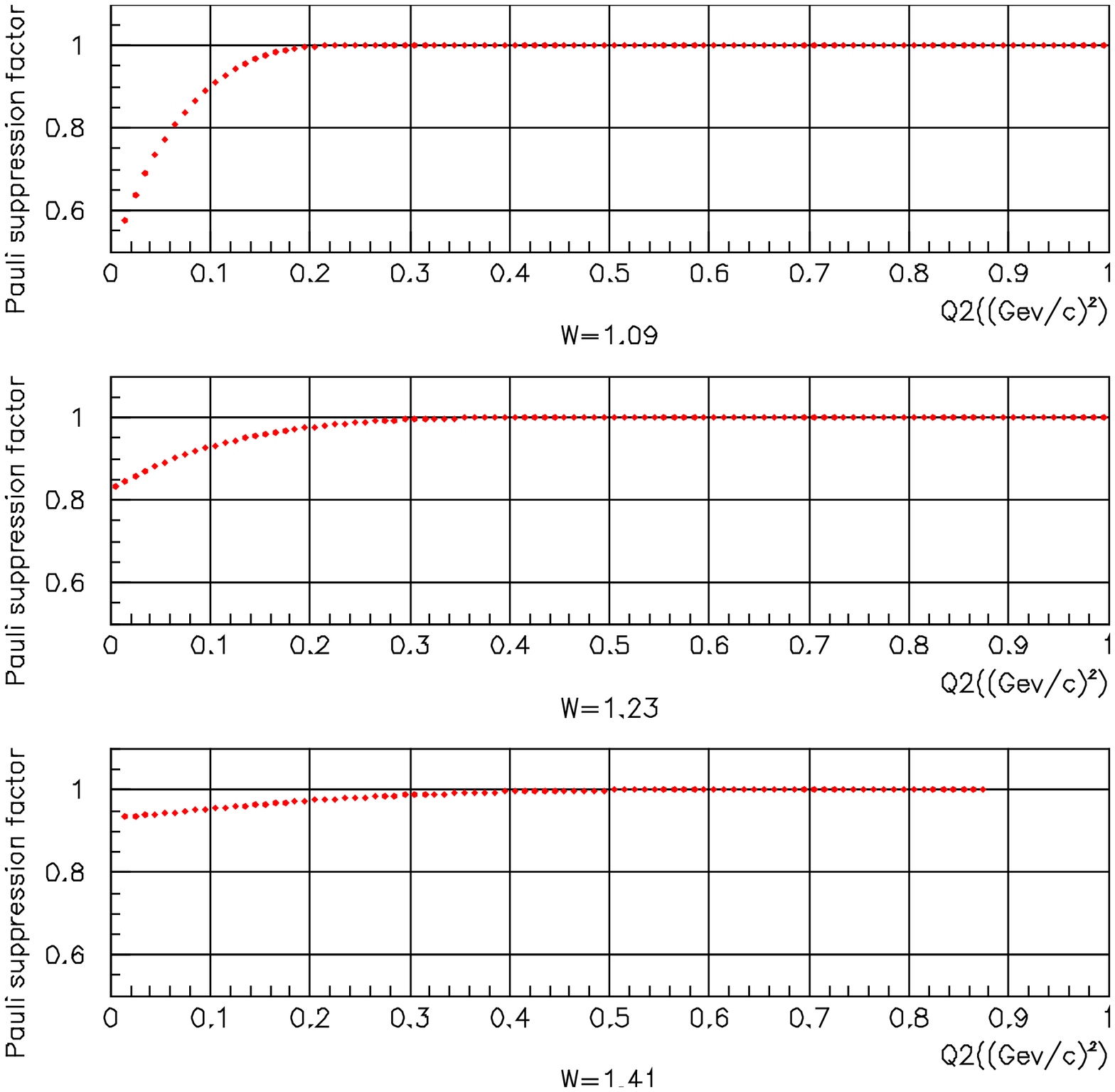}
%\subfigure{\epsfig{figure=pauli.eps,height=3.5in,angle=0}}
\vspace*{0.5cm}
\caption{\sf The Pauli suppression factor as a function of $Q^2$ 
and for various values of $W$, for the case of $P_F = 0.226\, {\rm GeV}/c$.}
\label{pausupp}
\end{figure}

\begin{figure}[htb]
\centering
\vspace*{-0.5cm}
\includegraphics[angle=0,width=7.5cm]{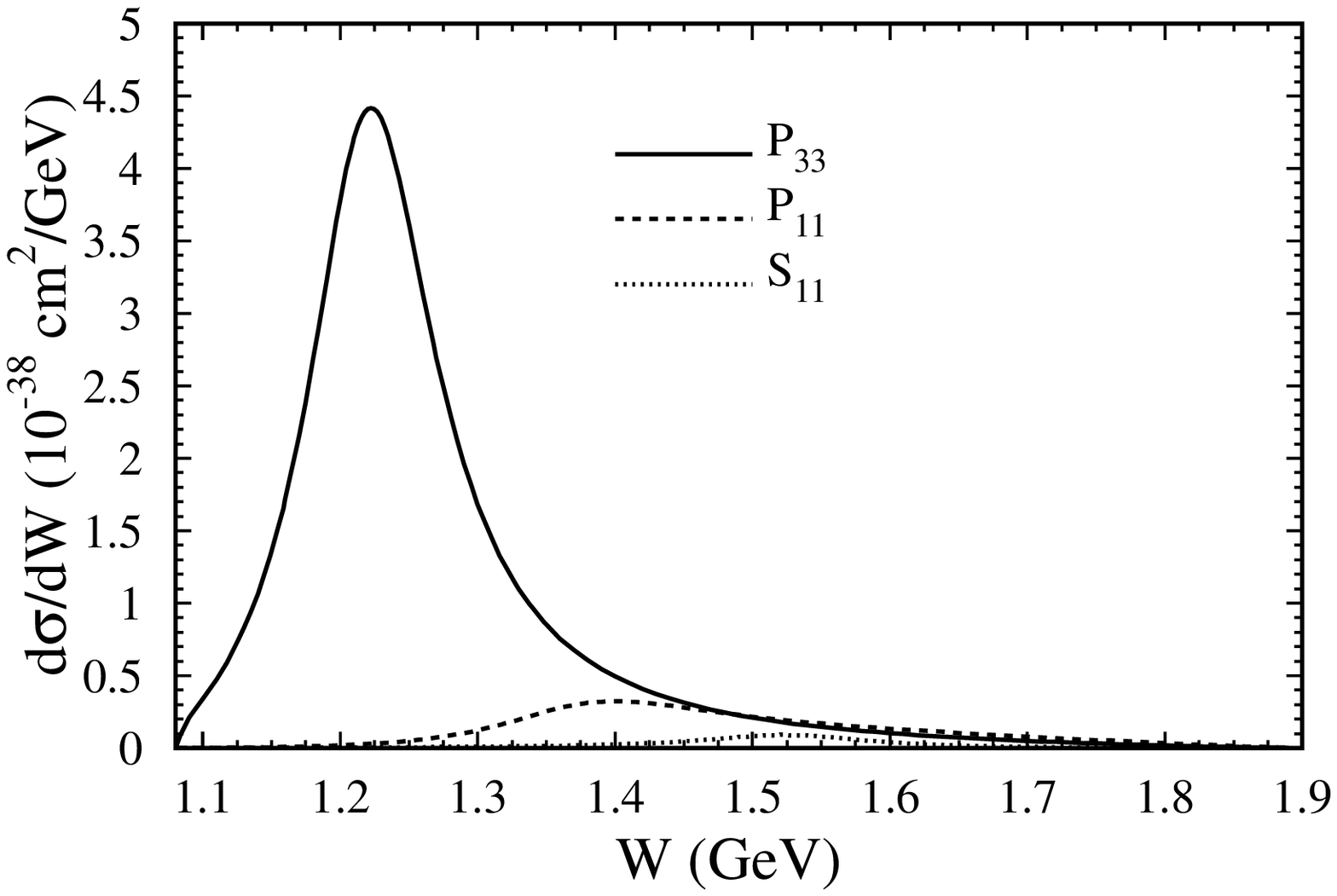}
\includegraphics[angle=0,width=7.5cm]{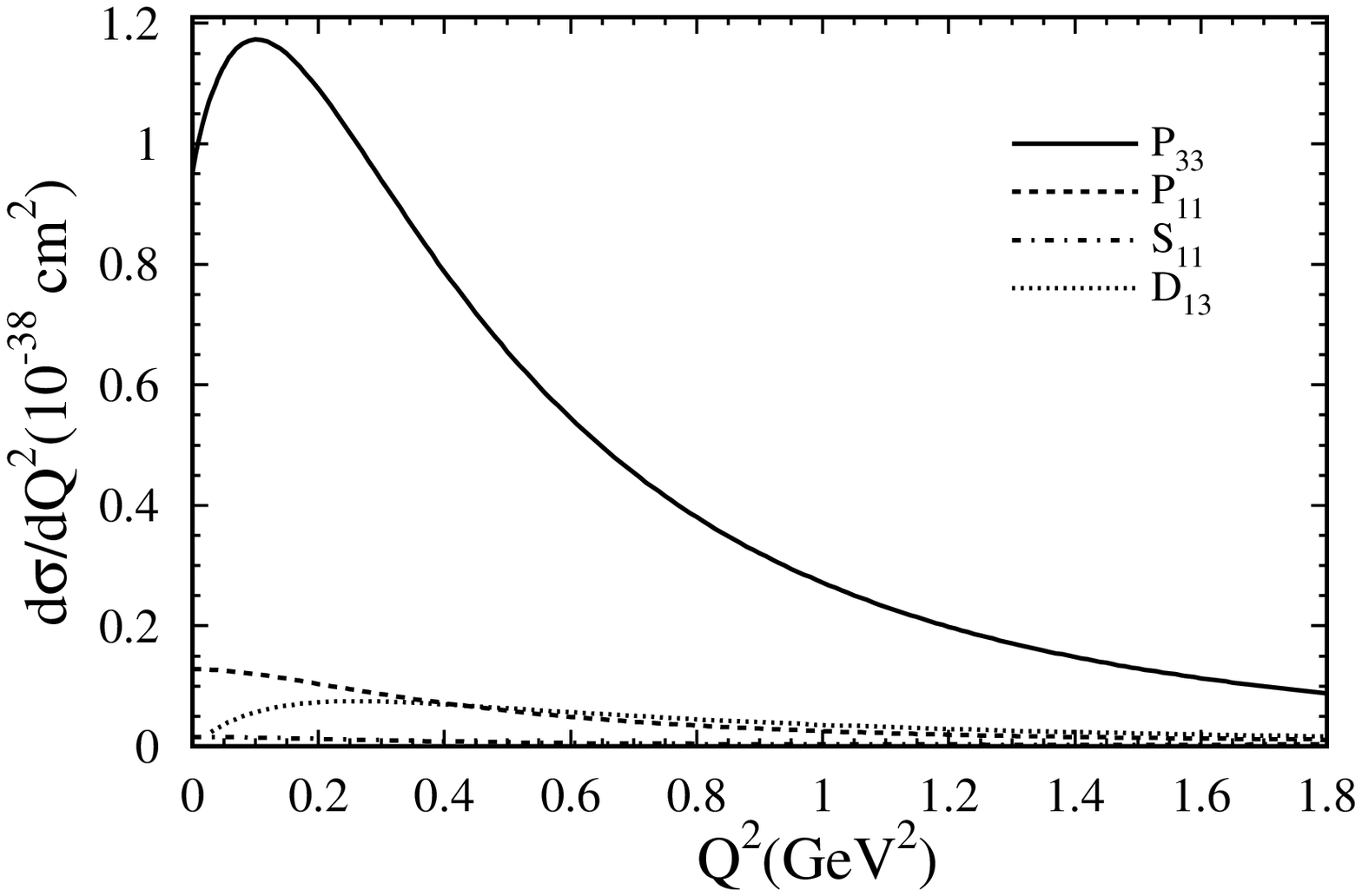}  
%\subfigure{\epsfig{figure=pauli.eps,height=3.5in,angle=0}}
\vspace*{-1.cm}
\caption{\sf Spectra of (a) invariant mass ${\rm d} \sigma/{\rm d} W$ 
and  (b) ${\rm d} \sigma/{\rm d} Q^2$ for $P_{33}, \,P_{11}$ 
and $S_{11}$ resonance with neutrino energy $E_\nu = 1.5 \, {\rm GeV}$. The curves correspond to each resonance alone without interferences.}
\label{dwres}
\end{figure}

For the production of the $\Delta$--resonance a similar
correction applies.  After the scattering of the 
current on a neutron the $\Delta$--resonance is
produced which travels and decays within the nucleus.
For most of the kinematic region the $\Delta$ is
non--relativistic and it takes 5 to 10 lifetimes to
travel through the nucleus.  When it decays it 
transfers momentum $|\vec{q}|$ to the proton
which is still bound.  For the $\Delta$--resonance
special attention is required since it propagates
in a medium.  However, once it decays we have a 
proton which seeks to find an empty state.  In the
independent particle of the Fermi model the 
unoccupied levels are above the Fermi surface.
The kinematics were considered analytically and
the blocking was given explicitly in ref. \cite{ref21}.
For completeness we transcribe the formulas here
introducing the notation familiar from deep inelastic
scattering.  We define in the rest frame of the
$\Delta$--resonance
\begin{displaymath}
p_{\pi} = \frac{W^2+m_{\pi}^2-M^2}{2W},\quad
q_0     = \frac{W^2-M^2-Q^2}{2W}\quad\quad{\rm{and}}\quad\quad R=p_F
\end{displaymath}
\begin{eqnarray}
{\rm (i)}&{\rm For}& \,2p_F\geq |\vec{q}|+p_{\pi}>|\vec{q}|-p_{\pi}\nonumber\\
&& g(W,|\vec{q}|) = \frac{1}{2|\vec{q}|}
\left( \frac{3|\vec{q}|^2+p_{\pi}^2}{2R} -
\frac{5|\vec{q}|^4+p_{\pi}^4+10|\vec{q}|^2q_{\pi}}{40R^3} \right)\\
{\rm (ii)}&& |\vec{q}|+p_{\pi}>2p_F \nonumber\\ 
&& g(W,|\vec{q}|) = \frac{1}{4p_{\pi}|\vec{q}|}
\left( (|\vec{q}|-p_{\pi})^2 -\frac{4}{5} R^2 -
\frac{(|\vec{q}|-p_{\pi})^3}{2R} +\frac
{(|\vec{q}|-p_{\pi})^5}{40R^3} \right)\\
{\rm (iii)}&& |\vec{q}|-p_{\pi}\geq 2p_F\nonumber\\
&& g(W,|\vec{q}|) = 1
\end{eqnarray}
The complicated formulas result from integrations
over the angles. For most of the phase space the
first term in the bracket is dominant.  We plot in
figure \ref{pausupp} the Pauli suppression factor as a 
function of $Q^2$ and for the various values of
$W$.  The suppression appears for $Q^2$
\raisebox{-0.1cm}{$\stackrel{<}{\sim}$} 0.2 GeV$^2$.

In the Monte--Carlo method, the Pauli exclusion
effect is taken into account by requiring the 
recoiling nucleon momentum to be greater than
$p_F$.  They obtain similar results \cite{ref22}.

The effects of Fermi motion are easily included.
Since the cross sections in eqs. (2.8) and (3.8) 
are written in a Lorentz invariant way, they are valid in any frame.
In the laboratory frame 
we give the proton a small momentum within its Fermi sea
\begin{eqnarray}
p_\mu = (\sqrt{p^2+M_N^2},p_x,p_y,p_z)
\end{eqnarray}  
and write the inner products $k\cdot p,\,k^\prime\cdot p$ accordingly.
Then one integrates numerically for all momenta inside the Fermi 
surface $|\vec{p}|<p_F$.

\setcounter{equation}{0}

\section{Numerical Results}

With the formalism described in sections 2, 3 and the Appendix A
we can study the contributions of the various terms to the cross sections.
We decided to include the resonances 
$\Delta(1232), P_{11}(1440), D_{13} (1520)$ and $S_{11}(1535)$.
The last two resonances are narrow and further way so that we expect a smaller 
contribution at lower energies.
For typical values of the parameters given in this article and using for 
$D_{13}(1520)$ couplings close to those of the $\Delta$-resonance, 
we computed the invariant mass and $Q^2$ contributions 
for two energies $E_\nu =1.5$ and $2.0\, {\rm GeV}$  
and found the results shown in figure 3. The curves show the cross section 
for $P_{33}$ going to 
$I =3/2$, and $S_{11}\ P_{11}$ separately going to the state with $I =1/2$.
The same situation prevails for $E_\nu =2.0\, {\rm GeV}$.
We can interpret this result as indicating the fact that the $P_{11}$ is 
closest to the $\Delta$-resonance and has a larger width.
For the relative low energies of the neutrino beams $E_\nu < 2.0 \,{\rm GeV}$ 
and $W < 1.4\, {\rm GeV}$ the dominant contribution comes from the $\Delta$-resonance
with an $I=1/2$ background from the other resonances 
and perhaps part of a continuum. 
The $I =1/2$ terms contribute only to the $p\pi^0$ and $n\pi^+$ final states.

\begin{figure}[htb]
\centering
\vspace*{-1.2cm}
\includegraphics[angle=0,width=9cm]{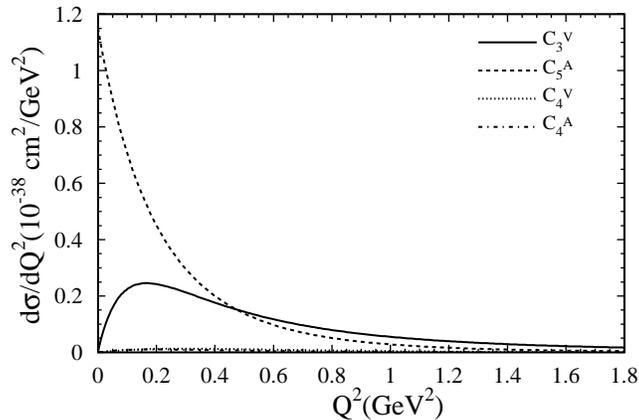}
%\subfigure{\epsfig{figure=pauli.eps,height=3.5in,angle=0}}
\vspace*{-1.5cm}
\caption{\sf Form factors}
\label{forf}
\end{figure}

\begin{figure}[htb]
\centering
\vspace*{-1.cm}
\includegraphics[angle=0,width=9cm]{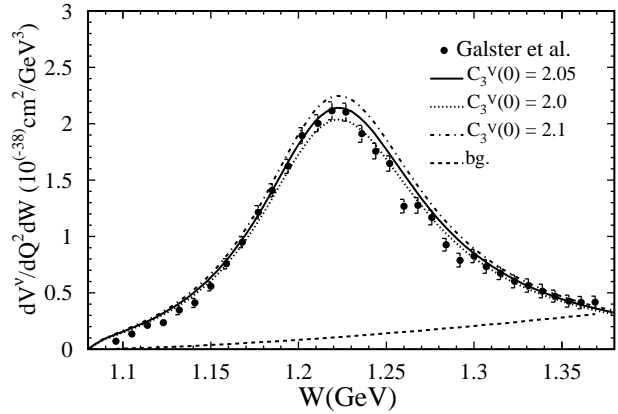}
%\includegraphics[angle=0,width=10cm]{bnl_daten.eps}
%\subfigure{\epsfig{figure=pauli.eps,height=3.5in,angle=0}}
\vspace*{-1.5cm}
\caption{\sf Cross section $dV^{\nu}/{dQ^2dW}$ for electroproduction in the $\Delta$ resonance.}
\label{wu0}
\end{figure}

As a next issue we consider the contribution of the various form factors. We 
show in figure 4 the relative importance of the various form factors, 
where $C_3^V$ and $C_5^A$ dominate the cross section.
The cross section from the axial form factors 
has a peak at $Q^2= 0$, while the cross section from $C_3^V$ turns to zero.
The zero from the vector
form factor is understood, because in the 
configuration where the muon is parallel to the neutrino, 
the leptonic current is 
proportional to $q_\mu$ and takes the divergence
of the vector current, which vanishes by CVC.  
The contributions from
$C_4^V$ and $C_4^A$ are very small as shown in
the figure \ref{forf}.
\vspace*{-0.3cm}
Thus the excitation of the $\Delta$-resonance, to the accuracy of present 
experiments, is described by two form factors.

An estimate of the vector contribution is also possible 
using electroproduction data.
There are precise data for the electroproduction of 
the $\Delta$ and other resonances \cite{ref11}
including their decays to various pion-nucleon modes.
In the data of Galster et al. cross sections for the channels $(p+\pi^0)$
and $(n+\pi^+)$ are tabulated from which we conclude that both $I=3/2$ 
and $I=1/2$ amplitudes are present. For instance, for $W = 1.232\, {\rm GeV}$ the 
$I = 1/2$ background is $10\%$ of the cross section.

For our comparison  we shall take the electroproduction 
data after subtraction of the background, as shown in figure 5, 
and then use CVC to obtain the contribution of $V_\mu^+$ to neutrino induced reactions.
We use the formula
\begin{equation}
\frac{dV^{\nu}}{dQ^2dW} = \frac{G^2}{\pi}\,\,
\frac{3}{8}\,\, \frac{Q^4}{\pi\alpha^2}\,\,
\frac{d\sigma^{{\rm em},\, I = 1}}{dQ^2dW}
\end{equation}
to convert the observed \cite{ref11} cross sections 
for the sum of the reactions
$e + p \rightarrow  e + \left\{\begin{array}{c} p \,\pi^0\\ n \,\pi^+ \end{array} \right. $ to the vector contribution in the reaction 
$\nu +p \rightarrow \mu^- + p + \pi^+$ denoted in eq. (5.1) by $V^{\nu}$.
The factor $3/8$ originates from the Clebsch-Gordan 
coefficients relating the matrix elements of the two 
channels in the electromagnetic case to the matrix 
element of the weak charged current.
We use the data of Galster et al. \cite{ref11} at $Q^2 = 0.35 \,{\rm GeV}^2$
and subtract the background as suggested by them.
Then we converted the points to the vector contribution 
for the neutrino reaction according to eq. (5.1).
In the same figure we show the 
neutrino cross section with $C_3^V(0) = 2.05$ (solid), 
$C_3^V(0) = 2.0$ (dotted), $C_3^V(0) = 2.1$ (dot-dashed) 
and contribution of background (dashed)   
and all other form factors equal to zero.
Before leaving this topic we mention that the analysis 
of the electroproduction data \cite{ref11} included a contribution from the 
$D_{13}(1520)$ resonance which was found to be small.

For the axial form factor we use the form given in eq. (2.6) 
and we must keep an open mind to notice 
whether a modification will become necessary.
With the method described in this article we have all parameters for the 
$\Delta$-resonance. We may still change the couplings by a 
few percent and vary $M_V$ and $M_A$. 
For the other resonances we can use the results
of section 3 which for the low energies introduce an $I =1/2$ background.
For higher energies and for other channels the additional resonances play 
significant role because they influence the overall scale of the 
integrated cross section; they also contribute to the multi-pion 
events since the inelasticities are large.

\begin{figure}[htb]
\centering
\vspace*{-1.cm}
\includegraphics[angle=0,width=9cm]{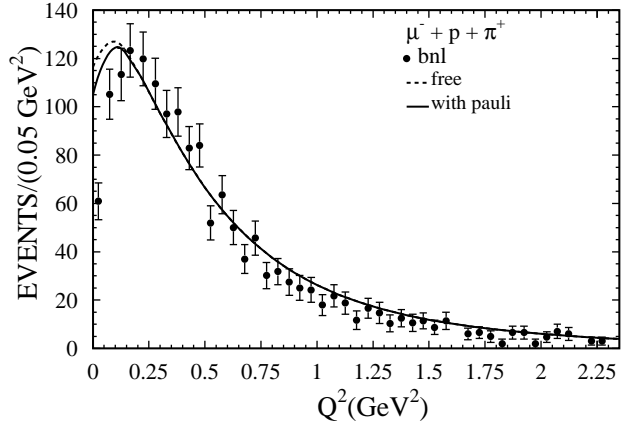}
%\includegraphics[angle=0,width=9cm]{bnl_daten.eps}
%\subfigure{\epsfig{figure=pauli.eps,height=3.5in,angle=0}}
\vspace*{-1.3cm}
\caption{\sf $Q^2$-spectrum of the process $\nu p \rightarrow \mu^- p \pi^+$.}
\label{bnl}
\end{figure}

There is still the $Q^2$ distribution \cite{ref16} 
to be accounted for. As mentioned already,
the data is from the Brookhaven experiment \cite{ref17} 
where the experimental group  
presented a histogram averaged over the neutrino flux 
and with an unspecified normalization.
We use the formalism of this article with the $\Delta$
resonances plus the correction from the Pauli factor.
For the relative normalization, we normalized the  area
under the theoretical curve for $Q^2 \geq 0.2 \,{\rm GeV}^2$ 
to the corresponding area under the data.
For the other parameters we choose
\begin{eqnarray}
&C_3^V(0) = 1.95\,&, \quad C_5^A(0) = 1.2\, ,\nonumber\\
&M_V = 0.84 \,{\rm GeV}\,&, \quad M_A = 1.05 \,{\rm GeV}\,.
\end{eqnarray}
The result is the solid curve and $P_F = 0.160 \, {\rm GeV}$ 
shown in figure \ref{bnl} which is satisfactory.
In fact we made a $\chi^2$-fit and obtained these values for $\chi^2$
equal to 1.76 per degree of freedom.
In the theoretical curves we averaged over the neutrino 
flux for the BNL experiment \cite{baker}.  
The dotted curve is the calculation without Pauli factor 
and the solid one with Pauli-effect.
This analysis will be repeated when new data becomes available.

Finally we calculated the integrated cross sections 
as function of the neutrino energy for the various channels. 
All cross sections reach at higher energies constant values.
The asymptotic value for the $p\pi^+$ channel depends 
on the excitation of the $\Delta$ and the input parameters.
For  the other channels, however, the shape of the integrated 
cross sections and the constant asymptotic value 
also depends on the $I=1/2$ contribution. 
In fact, comparisons of the available compilations with theoretical estimates 
show different values \cite{ref10,ref13}.
This topic will be investigated theoretically and in the new experiments, 
where additional contributions from inelastic channels 
and other nuclear effects must be considered.

\setcounter{equation}{0}

\section{Summary}

We adopted in this article a simplified formulation for the production 
of the $\Delta$-resonance which depends on two independent form factors.
One form factor for the vector current was determined 
from electroproduction data, and the other axial vector form factor 
determined by PCAC and neutrino data.
We plotted the vector and axial contributions separately 
in order to understand and locate their characteristic properties.
Our numerical results agree qualitatively with previous analysis \cite{ref8}.

We analyzed the differential cross section ${\rm d}\sigma/{\rm d} Q^2$ in
terms of the two form factors.
The vector contribution has a modified dipole form, 
as determined from electroproduction data and the axial form factor 
is also modified.
We find that the values at low- and high-$Q^2$ are correlated, 
since we had to search diligently for the values in eq. (5.2), which give
an acceptable fit of the data.
Fitting the large $Q^2$ region gives a  
diffential cross section which is too high at $Q^2\leq 0.15 \,{\rm GeV^2}$.
This may be due to the scanning efficiency \cite{Furuno} 
or may be a genuine property and must be followed up in the future. 
The present analysis points to the direction 
that a modified axial formfactor 
may be preferable.
Similar tendencies were reported for $Q^2$-distributions 
of other articles \cite{Furuno,bell,schreinerb}.

We also included the Pauli factor which 
brings a small correction at $Q^2\le 0.20 \, {\rm GeV}^2$.
In order to justify its application to light nuclei 
and for the decay of a resonance within a nucleus we rederived the Pauli factor
in the Fermi gas model and showed that it agrees with the standard geometrical 
interpretation.

%We analyzed the ${\rm d}\sigma/{\rm d} Q^2$ distribution 
%which was harder to fit with only two form factors.
%We found that there is a correlation between the high- and 
%low-$Q^2$ regions of the data. 
%This presented demands on the shape of the form factors 
%and the contribution of Pauli blocking.
%In fact after fitting the small $Q^2$ region we obtained 
%larger values at $Q^2\ge 1\, {\rm GeV}^2$.

%Among the nuclear correction we redrived a general formula 
%for Pauli suppression factor which has a known geometrical meaning.
%The Pauli effect helps in improving the $Q^2$-distribution.

Now, that the $\Delta^{++}$ resonance can be accounted 
for there is interest to predict the other channels 
$p\pi^0$ and $n\pi^+$, where $I=1/2$ resonances also contribute.
In section 3, we give formula for the differential cross 
section for $P_{11}$ and $S_{11}$ resonances. 
They will influence the cross sections for the other channel 
especially at energies $E_\nu >2 \, {\rm GeV}$.

In this article we have demonstrated that the excitation of the 
$\Delta$-resonance 
can be accounted for by two form factors.
This result forms the basis for analysis of new data 
which will confirm the adequacy of this minimal set 
or require additional form factors or alternative parameterizations.
The work will be extended to the other final states $p\pi^0$ and $n\pi^+$ 
where $I =1/2$ will be included. The extension 
to higher energies $E_\nu > 2\, {\rm GeV}$ will reveal the significance 
of additional resonances, especially those with large 
inelasticities, because they may reveal characteristics 
for the transition to the inelastic region where
multi-pion production is important.  

\pagebreak
\noindent{\large\bf{Acknowledgement}}

The support of the 
``Bundesministerium f\"ur Bildung, Wissenschaft, Forschung und
Technologie'', Bonn under contract 05HT1PEA9 is gratefully 
acknowledged.
We wish to thank Dr. I. Schienbein for helpful discussions 
and his interest on the connection between electroproduction 
and neutrino data, which he has also been investigating.
We also thank Prof. D. P. Roy 
and Mr. F. von Horsten for helpful discussions.
%\newpage

\subsection{\underline{Appendix A}:
Formulas for the amplitudes and the cross sections}

In this appendix we give a summary of the formulas
used for the calculation of the cross section.
We set the lepton mass equal to zero and write
the differential cross section as 
\begin{displaymath}
\frac{d\sigma}{dQ^2dW} =\frac{G^2}{16\pi M^2}\sum_{i=1}^3 K_i W_i.
\end{displaymath}
The kinematic factors are
\begin{eqnarray}
K_1 & = & \frac{2Q^2}{E_{\nu}^2}\nonumber\\
K_2 & = & 4\left( 1-\frac{Q^2}{4E_{\nu}^2}
          -\frac{q_L^0}{E_{\nu}}\right)\nonumber\\
K_3 & = & \frac{MQ^2}{E_{\nu}Wq_{CM}}
          \left(2-\frac{q_L^0}{E_{\nu}}\right)\nonumber\, 
\end{eqnarray} 
with the kinematic variables being those
used in deep inelastic scattering.

The structure functions $W_i$ are expressed in
terms of helicity amplitudes.
\begin{eqnarray}
W_1 & = & \frac{W}{q_{CM}}\left(|T_{\frac{3}{2}}|^2+ |T_{\frac{1}{2}}|^2
          + |U_{\frac{3}{2}}|^2 
          + U_{\frac{1}{2}}|^2\right)\nonumber\\
W_2 & = & \frac{M^2 Q^2}{W q_{CM}^3}
          \left(|T_{\frac{3}{2}}|^2+ |T_{\frac{1}{2}}|^2+
          |U_{\frac{3}{2}}|^2 + U_{\frac{1}{2}}|^2\right)
      +   \frac{2M^2Q^4}{W_{CM}^5}\left(|T_C|^2+|U_C|^2\right) \nonumber\\
W_3 & = & \frac{4W}{q_{CM}} \left( Re\, T_{\frac{3}{2}}^*\, U_{\frac{3}{2}} 
      -   Re\, T_{\frac{1}{2}}^* \, U_{\frac{1}{2}}\right)\nonumber
\end{eqnarray}

Several remarks are now in order.  The structure
functions $W_1$ and $W_2$ contain the square of
the vector current plus the square of the axial
currents.  The structure $W_3$ is the vector
$\otimes$ axial interference term.  The last term
in $W_2$ is regular as $Q^2\to 0$.  Finally, 
$T_i$ and $U_i$ are real so that the symbol for 
a real part is superfluous.

The helicity amplitudes are given in terms of
form factors
\begin{eqnarray}
T_{\frac{3}{2}} 
& = & y\left[ \left( \frac{W+M}{M} \right) C_3^V +\frac{Wq_{CM}^0}{M^2} 
      C_4^V\right]\nonumber\\
T_{\frac{1}{2}} 
& = & \frac{y}{\sqrt{3}}\left[ \frac{q_{CM}^0-p_{CM}^0-M}{M} C_3^V
  +   \frac{Wq_{CM}^0}{M_N^2} C_4^V\right]\nonumber\\
T_C 
& = & -\sqrt{\frac{2}{3}}y \frac{q_{CM}}{M}
       \left( C_3^V + \frac{W}{M} C_4^V \right)\nonumber.
\end{eqnarray}

The amplitude $T_C$ vanishes at $W=M_{\Delta}$.
Similarly the axial current contributions are
\begin{eqnarray}
U_{\frac{3}{2}} 
& = & z\left(\frac{Wq_{CM}^0}{M^2} C_4^A + C_5^A \right)\nonumber\\
U_{\frac{1}{2}} 
& = & \frac{1}{\sqrt{3}} U_{\frac{3}{2}}\nonumber\\
U_C 
& = & -\sqrt{\frac{3}{2}} z\frac{q_{CM}}{M}
       \left( \frac{W}{M} C_4^A -\frac{Mq_{CM}^0}{Q^2}C_5^A\right)\nonumber.
\end{eqnarray}
The kinematic factors are:
\begin{displaymath}
y =f(W)N_{RS}q_{CM}, \quad\quad
z = f(W)N_{RS}(p_{CM}^0+M),\, .
\end{displaymath}

\begin{displaymath}
N_{RS} = -i \left[ \frac{q_{CM}}{4W(p_{CM}^0+M)}\right]^{1/2},
\quad\quad
p_{CM}^0 = \left[q_{CM}^2+M^2\right]^{1/2}
\end{displaymath}
and the rest defined in section one.


\begin{thebibliography}{54}
\bibitem{ref20} The formula in eq.\ (3.33) 
        of ref.\ \cite{ref4} has a
        misprint where the third term 
        in the curly bracket, $\frac{x^3}{3}$, should be
        replaced by $\frac{x^3}{2}$.
\end{thebibliography}
\end{document}